\documentclass[final,1p,times]{elsarticle}
\usepackage{amsfonts}
\usepackage{amsmath}
\usepackage{geometry}
\newmuskip\pFqmuskip

\newcommand*\pFq[6][8]{%
  \begingroup 
  \pFqmuskip=#1mu\relax
  \mathchardef\normalcomma=\mathcode`,
  \mathcode`\,=\string"8000
  \begingroup\lccode`\~=`\,
  \lowercase{\endgroup\let~}\pFqcomma
  {}_{#2}F_{#3}{\left[\genfrac..{0pt}{}{#4}{#5};#6\right]}%
  \endgroup
}
\newcommand{\pFqcomma}{{\normalcomma}\mskip\pFqmuskip}
\usepackage{caption}
\usepackage{bm}

\newdefinition{rmk}{Remark}
\newproof{pf}{Proof}
\newproof{pot}{Proof of Theorem \ref{thm2}}
\newcommand{\be}{\begin{equation}}
\newcommand{\ee}{\end{equation}}
\newcommand{\bea}{\begin{eqnarray}}
\newcommand{\eea}{\end{eqnarray}}
\usepackage{fancyhdr} 
\journal{}
\begin{document}

\begin{frontmatter}
\title{Inexistence of quark mass in chiral symmetry and its relation to confinement dynamics}

\author{Yoon-Seok Choun}
\ead{ychoun@gmail.com}
\address{Department of Physics, Hanyang University, Seoul 04763, Korea}
\begin{abstract}
In 1985, G\"ursey showed that the spectrum of the semi-relativistic Hamiltonian for the bag model, introduced by Lichtenberg et al. for mesons, follows the Regge trajectory if the current quark mass is negligible. The model leads to the biconfluent Heun equation, which is a second-order linear ordinary differential equation (ODE) with a regular singularity at the origin and an irregular singularity at infinity. 
 Based on rigorous mathematical computation, it is concluded that the energy spectrum is consistent with the Regge trajectory only when the quark mass vanishes. From this result, we suggest that the chiral symmetry is a consequence of confinement dynamics.   
\end{abstract}

\begin{keyword}
Heun's equation, 3-term recurrence relation, Polynomial, Quark mass, Color confinement

\MSC{34L16,  33C47, 49K15}
\end{keyword}

\end{frontmatter}

\section{Introduction}
Calculation of Schr$\ddot{\mbox{o}}$dinger's equations for the harmonic oscillator, infinite square well, Morse, P\"oschl-Teller, pseudoharmonic oscillator, Coulomb and hyperbolic, non-central ring-shaped potential quantum systems, etc., involves application of mathematical and physical techniques such as the factorization method, Lie algebras, matrix elements, and quantum control \cite{Ushv1994,Dong2007}. 
A common feature of the above methods is that the second order linear ordinary differential equations (ODEs) given by the potentials lead us to a 2-term recurrence relation, when analyzed using a power series solution after factoring out the asymptotic behaviors of the wave equations and changing the independent variables. Hypergeometric-type functions consist of the 2-term recurrence relation and are easy to handle; however, wave functions having the 3-term recursive relation in their power series are very difficult to analyze mathematically. In this paper, we discuss the Schr$\ddot{\mbox{o}}$dinger equation having a quadratic potential with a Coulomb-like one. 

A radial wave function satisfies the equation
\begin{equation}
\left[-\frac{\hbar ^2}{2m} \left( \frac{d^2}{d{r}^2} + \frac{2}{r}\frac{d}{dr}\right) + \frac{\hbar ^2}{2m}\frac{L(L+1)}{r^2} + V(r)\right] R(r) = E
R(r)
\label{qq:1}
\end{equation}
with
\begin{equation}
V(r) = c r^2 + b r -\frac{a}{r} \label{cornell}
\end{equation}
where $0\leq r < \infty $, $E$ is the eigenvalue, $L$ is the rotational quantum number, $c > 0$ and $a,b \in \mathbb{R}$.

Eq.(\ref{cornell}), consisting of a Coulombic term and a quark confining potential, is perceived in quark dynamics in relation to the concepts of asymptotic freedom and quark confinement in non-Abelian gauge theories such as non-relativistic quark-antiquark bound states described by a Schr\"odinger equation \cite{Appe1975,Eich1976,Eich1978}. Eq.(\ref{cornell}) was also studied by Gupta and Khare \cite{Gupt1977}.

A solution of the form $R(r) = r^{L} f(\tilde{r})$ is obtained by putting $\tilde{r} = r/\alpha $ and $\alpha = \frac{\hbar ^2}{2m}$. Eq.(\ref{qq:1}) then becomes  
\begin{equation}
 \frac{d^2 f(\tilde{r})}{d{\tilde{r}}^2} + \frac{2(L+1)}{\tilde{r}} \frac{d f(\tilde{r})}{d\tilde{r}} + \left( \tilde{E} -\tilde{c}^2 \tilde{r}^2- \tilde{b} \tilde{r}
 + \frac{\tilde{a}}{\tilde{r}} \right) f(\tilde{r}) = 0
\label{qq:4}
\end{equation}
where $\tilde{E} = \alpha E $, $\tilde{c}^2 = \alpha^3 c$, $\tilde{b} = \alpha^2 b$, and $\tilde{a} = a$.
If one requires solutions of the form
$ f(\tilde{r}) = \exp\left( - \frac{\tilde{c}}{2}\tilde{r}^2 -\frac{\tilde{b}}{2\tilde{c}}\tilde{r}\right)y(\tilde{r})$ in Eq.(\ref{qq:4}), then substituting $\rho = \sqrt{\tilde{c}}\tilde{r}$ into the new equation (\ref{qq:4}),
we obtain
\begin{footnotesize}
\begin{equation}
 \rho \frac{d^2 y(\rho)}{d\rho^2} + \left(-2\rho^2 -\frac{\tilde{b}}{\tilde{c}^{3/2}}\rho + 2(L+1) \right) \frac{d y(\rho)}{d\rho} + \left(
 \frac{1}{\tilde{c}}\left(\tilde{E} + \frac{\tilde{b}^2}{4\tilde{c}^2} -(2L+3)\tilde{c}\right) \rho + \frac{\tilde{a}}{\sqrt{\tilde{c}}}
 - \frac{\tilde{b}(L+1)}{\tilde{c}^{3/2}} \right) y(\rho) = 0
 \label{qq:6}
\end{equation}
\end{footnotesize}
Comparing Eq.(\ref{qq:6}) with Eq.(\ref{eq:1}), it is seen that the former is a special case of the latter with $z = \rho $, $\mu = -2$, $\varepsilon = -\frac{\tilde{b}}{\tilde{c}^{3/2}}$, $\nu  =  2(L+1)$, $\omega = L + 1 -\frac{\tilde{a}\tilde{c}}{\tilde{b}}$, and $\Omega  = \frac{1}{\tilde{c}}\left( \tilde{E} + \frac{\tilde{b}^2}{4\tilde{c}^2} -(2L+3)\tilde{c}\right)$.

We investigate the asymptotic behavior of the radial wave function $R(r)$ in Eq.(\ref{qq:1}) as the variable $r = \alpha \tilde{r}$ approaches positive infinity. We assume that $y(\rho)$ is an infinite series in Eq.(\ref{qq:6}) and substitute Eq.(\ref{eq:19}) in $R(r) = r^L \exp\left( -\frac{\tilde{c}}{2}\tilde{r}^2-\frac{\tilde{b}}{2\tilde{c}}\tilde{r}\right) y(\rho)$
 \begin{eqnarray}
R(r) &\sim &  \mathcal{A}\; \left( \alpha \tilde{r} \right)^L \exp\left( -\frac{\tilde{c}}{2}\tilde{r}^2-\frac{\tilde{b}}{2\tilde{c}}\tilde{r}\right) \tilde{z}^{\frac{\Omega}{2\mu}-\gamma}\exp(\tilde{z}) \nonumber\\ 
&= & \mathcal{A}\; r^L \left( \sqrt{\frac{c}{\alpha }} r^2 \right)^{\frac{-1}{4\tilde{c}}\left( \tilde{E} +\frac{\tilde{b}^2}{4\tilde{c}^2} \right) -\frac{1}{2}\left( L+\frac{3}{2} \right)} \exp\left(  \frac{1}{2}\sqrt{\frac{c}{\alpha }} r^2-\frac{b}{2}\sqrt{\frac{\alpha}{c}} r\right) \label{eq:20}
\end{eqnarray}
In Eq.(\ref{eq:20}), if $r\rightarrow \infty $, then $ R(r) \rightarrow \infty $. It is unacceptable for the wave function $R(r)$ to be divergent as $r$ approaches infinity, from the quantum mechanical point of view. Therefore, $y(\rho)$ must be a polynomial in Eq.(\ref{qq:6}) to render $R(r)$ convergent, even if $r$ approaches infinity. 

\section{Quantization of $\tilde{c}$ and the shooting method for BCH polynomials}
It was believed that the normalizable wave function is obtained by tuning the energy eigenvalue, irrespective of the form of the Schr$\ddot{\mbox{o}}$dinger equation. However, this is not possible for the biconfluent Heun (BCH) equation. Another parameter, besides energy, is required to build the polynomial solution for the Heun equations and their confluent forms \cite{Down2013,Down2016,Turb1988} as their series expansions consist of a 3-term recurrence relation. For example, $\tilde{c}$ ($\tilde{b}$ or $\tilde{a}$) and $\tilde{E}$ in Eq.(\ref{qq:6}). Meanwhile, hypergeometric-type functions are composed of a 2-term recursive relation; thus, we can construct a normalizable polynomial solution by tuning the single parameter of energy. The necessary and sufficient condition for constructing polynomials with a single parameter (the energy eigenvalue) is that their power series must be reduced to the 2-term recurrence relation, which is not possible for the Heun case, including its confluent forms. We demonstrate the reason why polynomials of the BCH equation cannot be described with a single parameter, based on which we build polynomials with two parameters, namely, $\tilde{c}$ and $\tilde{E}$. 

To obtain the polynomials of Eq.(\ref{qq:6}) around $\rho = 0$, we consider $\tilde{a}$, $\tilde{b}$ to be free variables, $-\Omega /\mu = \frac{1}{2\tilde{c}}\left( \tilde{E} + \frac{\tilde{b}^2}{4\tilde{c}^2} -(2L+3)\tilde{c}\right)$ to be a positive integer, and $\tilde{c}$ to be a fixed value.
From Eq.(\ref{eq:3}), we can see that a series expansion such as Eq.(\ref{eq:2}) becomes a polynomial of degree $N$ by imposing two conditions of the form \cite{Ronv1995}
\begin{equation}
B_{N+1} = d_{N+1} = 0 \hspace{1cm}\mathrm{where}\;N\in \mathbb{N}_{0}
 \label{eq:21}
\end{equation}
Eq.(\ref{eq:21}) is sufficient to yield $d_{N+2} = d_{N+3} = d_{N+4} = \cdots = 0$ successively and the solution of Eq.(\ref{qq:6}) becomes a polynomial of order $N$.

 The general expression of a power series of Eq.(\ref{qq:6}) about $\rho = 0$ for the polynomial and its algebraic equation for the determination of an accessory parameter $\tilde{c}$ are given by
(i) For $N = 0$,  Eq.(\ref{eq:21}) gives $B_1 = \frac{-\Omega}{2(2L+3)} = 0$ and $d_1 = A_0 d_0 = \frac{\frac{\tilde{b}}{\tilde{c}^{3/2}}(L+1-\frac{\tilde{a}\tilde{c}}{\tilde{b}})}{2(L+1)} d_0 = 0$.  If we choose $d_{0} = 0$, the entire series solution
vanishes. Therefore, $\tilde{c} = \frac{\tilde{b}(L+1)}{\tilde{a}}$ with $\tilde{E} = \frac{\tilde{b}}{\tilde{a}}(L+1)(2L+3)-\frac{\tilde{a}^2}{4(L+1)^2}$ at $L = 0,1,2,\cdots$. Its eigenfunction is $y(\rho) = \sum_{n = 0}^{0}d_n \rho^n = 1$, where $d_0 = 1$, for simplicity.

(ii) For $N = 1$, $B_2 = \frac{-\Omega + 2}{3(2L+4)}$ and $d_2 = A_1 d_1 + B_1  =  A_0 A_1 + B_1  =   \frac{\tilde{b}^2}{\tilde{c}^3} \frac{(L+1-\frac{\tilde{a}\tilde{c}}{\tilde{b}})(L+2-\frac{\tilde{a}\tilde{c}}{\tilde{b}})}{2(2L+2)(2L+3)} - \frac{\Omega }{2(2L+3)} $. Considering both $B_{2}$ and $d_{2}$ to be zero, we obtain $\tilde{E} = (2L+5)\tilde{c}-\frac{\tilde{b}^2}{4\tilde{c}^2} $ with $L = 0,1,2,\cdots$, and $\tilde{c}$ is given by the roots of a quadratic equation such as $2(2L+2)\tilde{c}^3-\tilde{a}^2 \tilde{c}^2 + \tilde{a}\tilde{b}(2L+3)\tilde{c}-\tilde{b}^2(L+1)(L+2) = 0$.
In this case, $y(\rho) = \sum_{n = 0}^{1}d_n \rho^n = 1 - \frac{\tilde{b}(L+1-
\frac{\tilde{a}\tilde{c}}{\tilde{b}})}{2(L+1)\tilde{c}^{3/2}} \rho$. 

(iii) For $N\geq2$, the eigenvalue is obtained by putting $B_{N+1} = 0$ and we get $\tilde{E} = 2\tilde{c} \left( N+L+\frac{3}{2}\right) -\frac{\tilde{b}^2}{4 \tilde{c}^2} $, where $N \in \mathbb{N}_{0}$ and $L = 0,1,2,\cdots, N$. The roots of $\tilde{c}$'s are obtained by putting $d_{N+1} = 0$, and its eigenfunction is $\sum_{n=0}^{N}d_n \rho^n$. (We are only interested in real roots owing to the physical conditions.) 

The ground state energy and its eigenfunction with a quantized $\tilde{c}$, where $N = 0$ and $L = 0$, are given by
\begin{footnotesize}
\begin{equation}
\begin{cases}
\tilde{E} = 3\tilde{c} -\frac{\tilde{b}^2}{4 \tilde{c}^2} \cr
\tilde{c} = \frac{\tilde{b}}{\tilde{a}} \cr
R(r) = \exp\left( -\frac{1}{2}\tilde{c}\tilde{r}^2-\frac{\tilde{b}}{2\tilde{c}}\tilde{r}\right)
\end{cases}
\label{eq:33}
\end{equation}
\end{footnotesize} 

We now demonstrate that polynomials of the BCH equation cannot be described with a single parameter $\widetilde{E}$ by applying the shooting method in general.
For $\tilde{a} = 2/5$ and $\tilde{b} = 1$ in Eq.(\ref{eq:33}), $\widetilde{E} = E_0 = 7.46$ and $\tilde{c} = c_0 = 2.5$ for the ground state and its polynomial is unity.

We assume that we can construct a polynomial of Eq.(\ref{qq:6}) for the ground state with only a single variable $\widetilde{E}$ without quantizing $\tilde{c}$. Then, any value of $\tilde{c}$ will be a fixed value $\tilde{c}$ for the polynomial of the equation. For $\tilde{c} = c_0 + 1.0$, we try to obtain a suitable value of $\widetilde{E}$ with initial conditions $y(0) = d_0 = 1$ and $y'(0) = 0$. Then, we construct a normalizable solution such as $y(\rho ) = 1$ by using the shooting method.

\begin{table}[!htb]
\footnotesize
    \begin{minipage}{.5\linewidth}
      \centering
        \begin{tabular}{|l|l|}
\hline
(1) & $\widetilde{E}= E_0-0.19$                         \\ \hline
(2) & $\widetilde{E}= E_0-0.15$                      \\ \hline
(3) & $\widetilde{E}= E_0-0.18$           \\ \hline
(4) & $\widetilde{E}= E_0-0.16$                  \\ \hline
(5) & $\widetilde{E}= E_0-0.17553$                    \\ \hline
(6) & $\widetilde{E}= E_0-0.17552$   \\ \hline
(7) & $\widetilde{E}= E_0-0.1755298911$   \\ \hline
(8) & $\widetilde{E}= E_0-0.1755298910$   \\ \hline
(9) & $\widetilde{E}= E_0-0.175529891060062$   \\ \hline
(10) & $\widetilde{E}= E_0-0.175529891060061$   \\ \hline
(11) & $\widetilde{E}= E_0-0.17552989106006135211$   \\ \hline
(12) & $\widetilde{E}= E_0-0.1755298910600613521149071$ \\ \hline
\end{tabular}
\caption{\scriptsize $\widetilde{E}$ of $y(\rho)$ where $\tilde{c} = c_0 + 1.0$.\\ See Fig. \ref{One}}
    \end{minipage}%
    \begin{minipage}{.5\linewidth}
      \centering
        \begin{tabular}{|l|l|}
\hline
(1)  & $\widetilde{E}=7.45$  \\ \hline
(2)  & $\widetilde{E}=7.461$   \\ \hline
(3)  & $\widetilde{E}=7.459$ \\ \hline
(4)  & $\widetilde{E}=7.4601$   \\ \hline
(5)  & $\widetilde{E}=7.4599$   \\ \hline
(6)  & $\widetilde{E}=7.46001$   \\ \hline
(7)  & $\widetilde{E}=7.45999$  \\ \hline
(8)  & $\widetilde{E}=7.460001$   \\ \hline
(9)  & $\widetilde{E}=7.459999$  \\ \hline
(10) & $\widetilde{E}=7.4600001$   \\ \hline
(11) & $\widetilde{E}=7.4599999$  \\ \hline
\end{tabular}
 \caption{\scriptsize $\widetilde{E}$ of $y(\rho)$ where $\tilde{c} = c_0$.\\ See Fig. \ref{Two}}
    \end{minipage}
\end{table}

\begin{figure}[!htb]
\minipage{0.5\textwidth}
  \includegraphics[width = \linewidth]{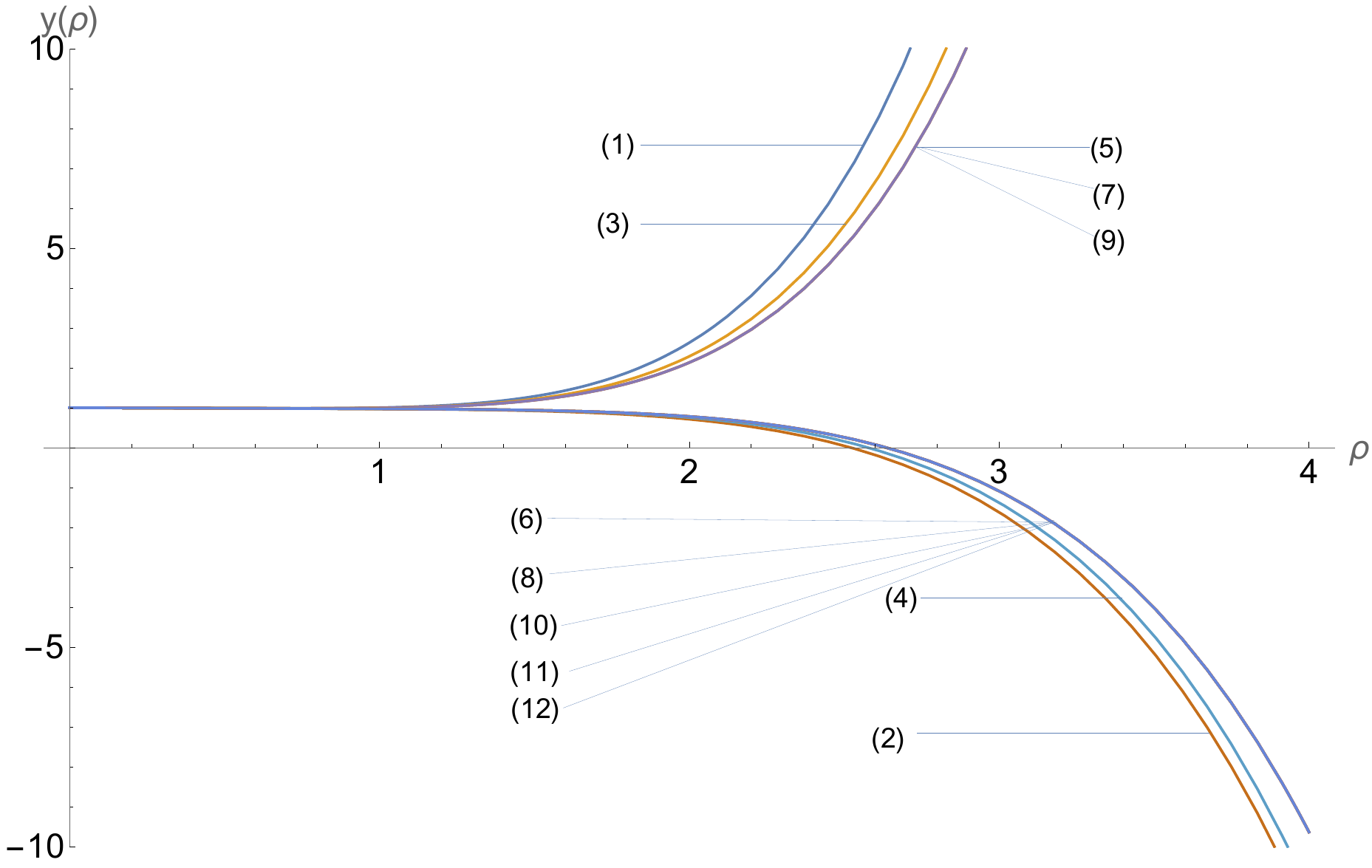}
  \caption{ $y(\rho)$ with a fixed $\tilde{c} = c_0 + 1.0$ and unfixed $\widetilde{E}$'s \\as  $\tilde{a} = 2/5$ and $\tilde{b} = 1$}
\label{One}
\endminipage\hfill
\minipage{0.5\textwidth}
 \includegraphics[width = \linewidth]{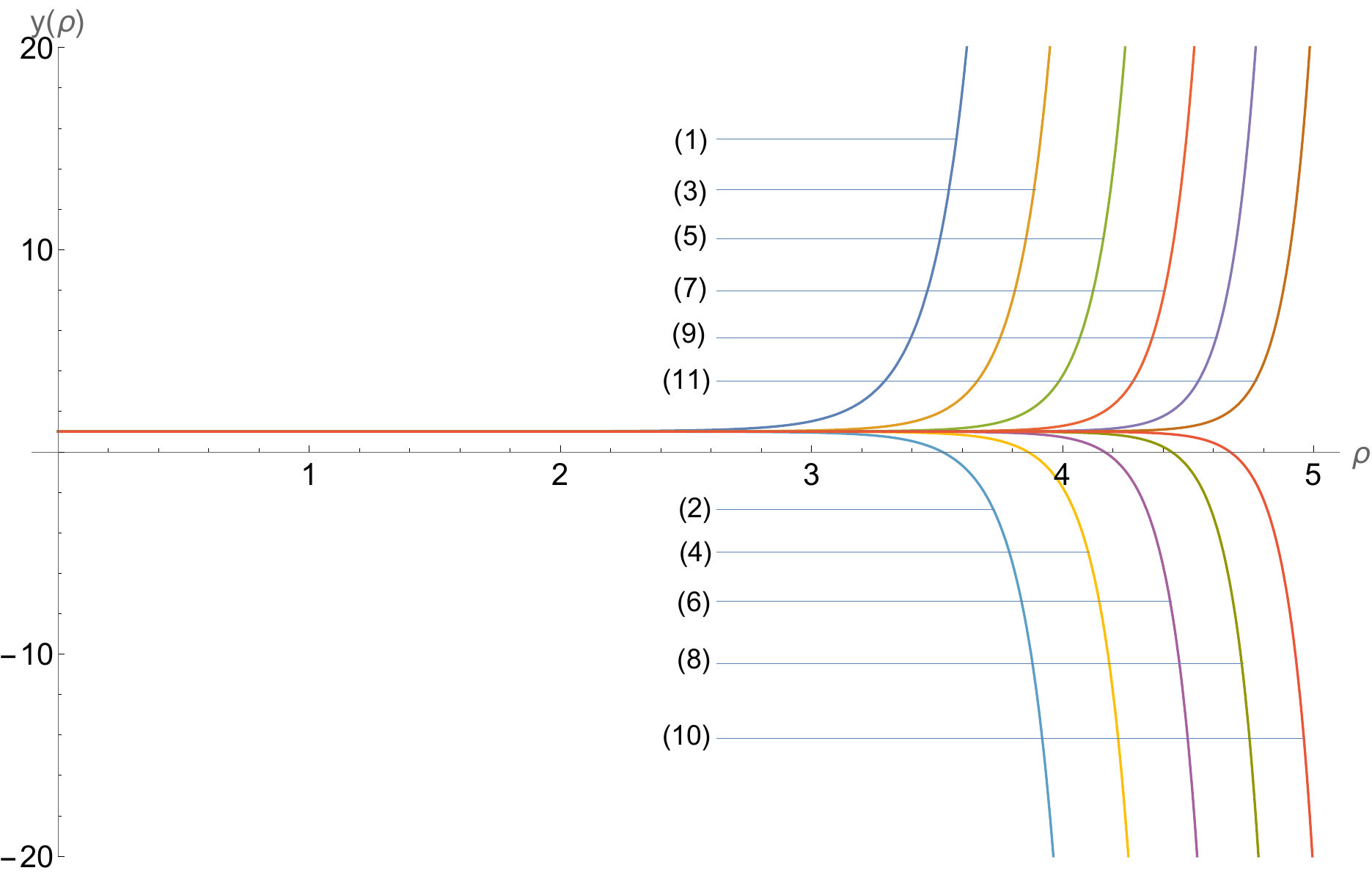}
  \caption{ $y(\rho)$ with a fixed $\tilde{c} = c_0$ and unfixed $\widetilde{E}$'s as  $\tilde{a} = 2/5$ and $\tilde{b} = 1$}
\label{Two}
\endminipage
\end{figure}

Figure \ref{One} shows how the trial wave functions approach unity
as we increase the precision of the eigenvalue $\widetilde{E}$. The functions (1), (3), (5), (7), and (9) are the under-shooted solutions and the functions (2), (4), (6), (8), (10), (11), and (12) are the over-shooted ones.
Starting from an under-shooted solution (1) at $\widetilde{E} = E_0 - 0.19$, we can increase the precision of $\widetilde{E}$ by increasing the minimal amount in the next digit to
obtain the over-shooted solution. Again, starting from an over-shooted solution (2) at $\widetilde{E} = E_0 - 0.15$, we can increase the precision of $\widetilde{E}$ by decreasing the minimal amount in the next digit to obtain the under-shooted solution.
After a number of iterations, the solutions cease to approach unity although we increase the precision by alternating the process of over- and under-shooting.
We observe that there is a limit to pushing the value of the function to the right as can be seen from the overlapped functions (5), (7), (9), and (6), (8), (10), (11), and (12).
When $\widetilde{E}$ reaches a value of nearly $ E_0 - 0.17553$, the functions start to flip drastically without being pushed to the right any further.

It can be seen from Fig.~\ref{Two} that $y(\rho)$ is pushed more to the right as $\widetilde{E} $ approaches 7.46 with $\tilde{c} = 2.5$ and if $\widetilde{E} $ is exactly equal to 7.46, $y(\rho) = 1$.
This situation occurs because a series expansion of Eq.(\ref{qq:6}) consists of a 3-term recurrence relation, which requires two quantized parameters ($\widetilde{E} $ and $\tilde{c}$) to create a polynomial.

If $\tilde{a} = \tilde{b} = 0$ in Eq.(\ref{qq:6}), by putting $\xi = \rho^2$, we obtain
\begin{equation}
 \xi \frac{d^2 y(\xi)}{d\xi^2} + \left(L+\frac{3}{2} - \xi \right) \frac{d y(\xi)}{d\xi} + \left( \frac{\widetilde{E}}{4\tilde{c}} - \frac{2L+3}{4} \right)
 y(\xi) = 0
\label{eq:41}
\end{equation}
Eq.(\ref{eq:41}) is a confluent hypergeometric equation and its series expansion is a 2-term recurrence relation. Its eigenvalue is $\widetilde{E} = 2\tilde{c}\left( N + L + 3/2\right)$ where $N = 0,1,2,\cdots$.
We observe that $\tilde{c}$ is not a fixed value any more but a free variable.  
By applying the shooting method with initial conditions $y'(0) = 0$ and $y(0) = 1$ in Eq.(\ref{eq:41}),we can obtain a suitable numerical value of $\widetilde{E}$.

\begin{table}[!htb]
\footnotesize
    \begin{minipage}{.5\linewidth}
      \centering
        \begin{tabular}{|l|l|}
\hline
(1) & $\widetilde{E}= 7.4 $                         \\ \hline
(2) & $\widetilde{E}= 7.50$                      \\ \hline
(3) & $\widetilde{E}= 7.49$           \\ \hline
(4) & $\widetilde{E}= 7.501$                  \\ \hline
(5) & $\widetilde{E}= 7.499 $                    \\ \hline
(6) & $\widetilde{E}= 7.5001$   \\ \hline
(7) & $\widetilde{E}= 7.4999 $   \\ \hline
(8) & $\widetilde{E}= 7.50001 $   \\ \hline
\end{tabular}
\caption{\scriptsize $\widetilde{E}$ of $y(\xi)$ where $\tilde{c} = c_0 = 2.5$.\\ See Fig.~\ref{Three}}
    \end{minipage}%
    \begin{minipage}{.5\linewidth}
      \centering
        \begin{tabular}{|l|l|}
\hline
(1)  & $\widetilde{E}= 10.4$  \\ \hline
(2)  & $\widetilde{E}= 10.6$   \\ \hline
(3)  & $\widetilde{E}= 10.49$ \\ \hline
(4)  & $\widetilde{E}=10.51 $   \\ \hline
(5)  & $\widetilde{E}= 10.499$   \\ \hline
(6)  & $\widetilde{E}=10.501 $   \\ \hline
(7)  & $\widetilde{E}=10.5000 $  \\ \hline
\end{tabular}
 \caption{\scriptsize $\widetilde{E}$ of $y(\xi)$ where $\tilde{c} = c_0 + 1.0 = 2.6$.\\ See Fig.~\ref{Four}}
    \end{minipage}
\end{table}

\begin{figure}[!htb]
\minipage{0.5\textwidth}
  \includegraphics[width = \linewidth]{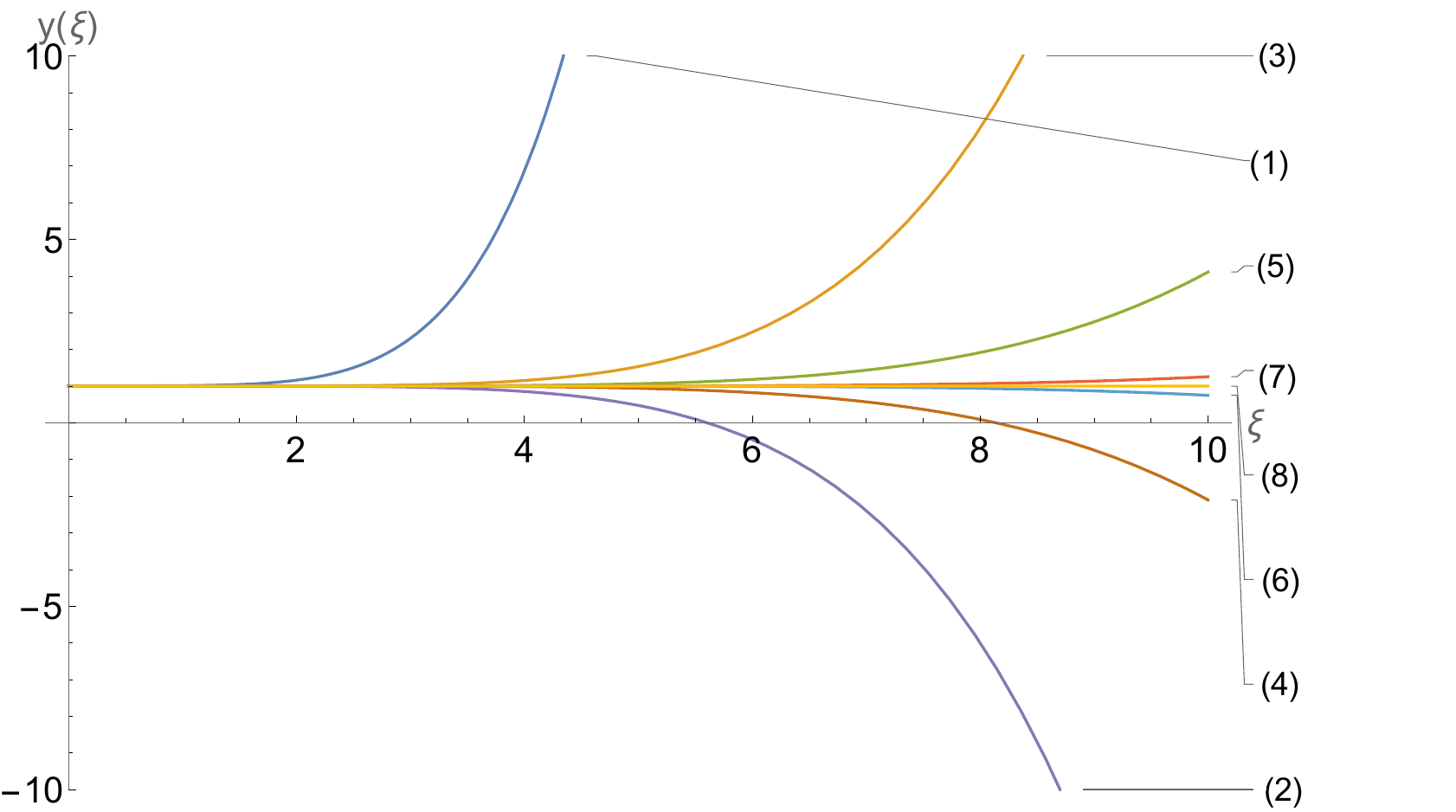}
  \caption{ $y(\xi)$ with a fixed $\tilde{c} = c_0 = 2.5$ and unfixed $\widetilde{E}$'s \\as $\tilde{a} = \tilde{b} = 0$}
\label{Three}
\endminipage\hfill
\minipage{0.5\textwidth}
 \includegraphics[width = \linewidth]{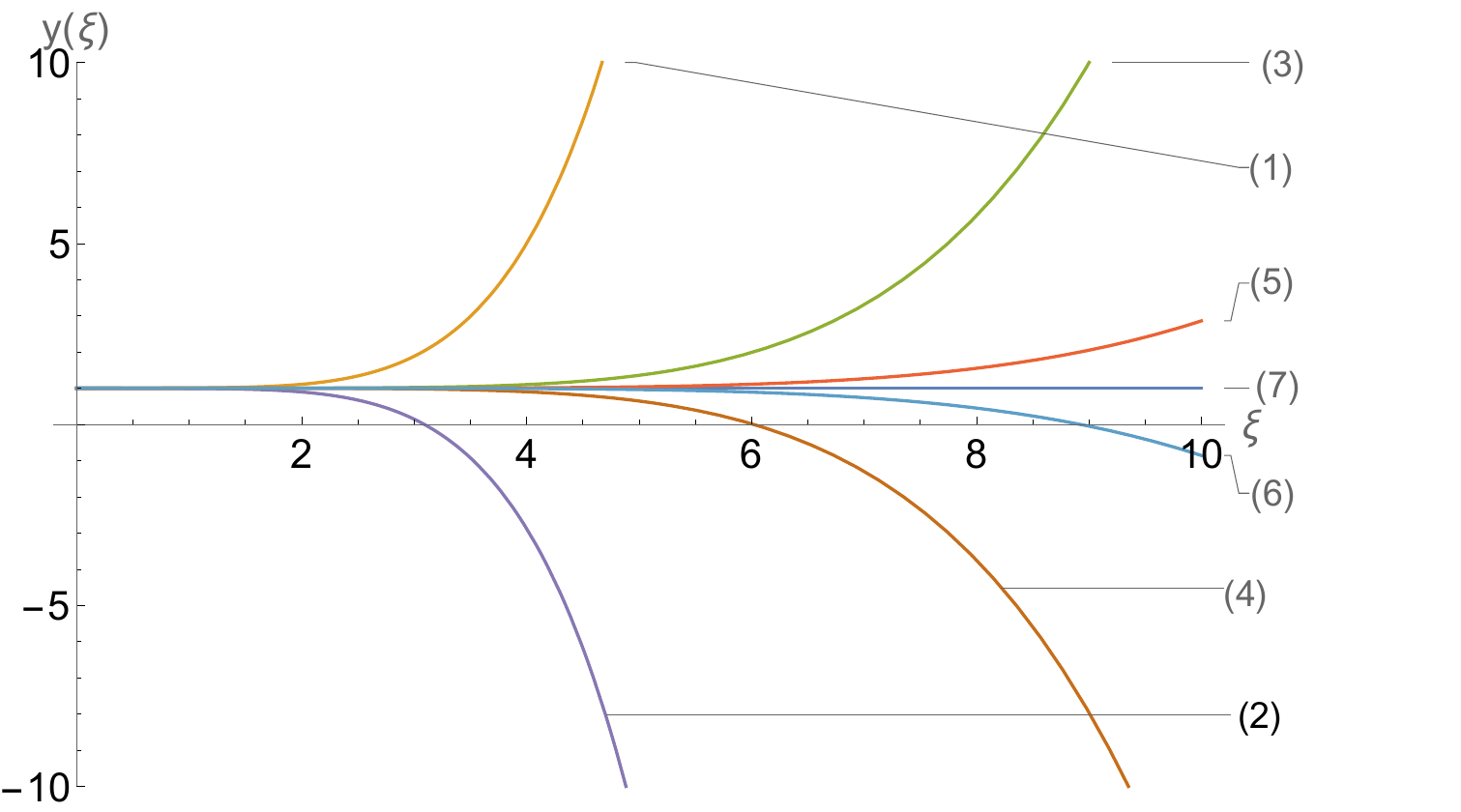}
  \caption{ $y(\xi)$ with a fixed $\tilde{c} = c_0 + 1.0$ and unfixed $\widetilde{E}$'s \\as $\tilde{a} = \tilde{b} = 0$}
\label{Four}
\endminipage
\end{figure}

Figure~\ref{Three} shows that $y(\xi)$ approaches unity for the ground state with $\tilde{c} = c_0 = 2.5$ as $\widetilde{E}$ reaches 7.5.
In Fig.~\ref{Four}, we observe that $y(\xi)$ approaches unity for the ground state for $\tilde{c} = c_0 + 1.0$ as $\widetilde{E}$ reaches 10.5. These examples demonstrate that two parameters, namely, ($\widetilde{E}$ and $\tilde{c}$) are required to obtain a polynomial
when a power series solution has a 3-term recurrence relation.
However, we need a single parameter, ($\widetilde{E}$), to obtain a polynomial when a series solution consists of a 2-term recursion relation.

\section{Application: Quark-antiquark system with scalar interaction }
In 1974, Wilson showed that the string theory introduced an approximation in the strongly interacting regime of QCD with the lattice gauge theory and proposed a computable strong-coupling limit of QCD \cite{Wils1974}.
In 1975, Eguchi demonstrated that two quarks squeezed together and formed a bilocal linear structure with a quark at one end and a diquark at the other end, through the string approximation with high rotational excitation \cite{Eguc1975}. A year later, Johnson and Thorn, following the bag model for a baryon structure, showed through QCD \cite{Chod1974.a,Chod1974.b} that the elongated bag model, whose structure is controlled by tubes of color flux lines stretched in a rotationally excited baryon, had a nearly linear Regge trajectory; they calculated the universal Regge slope $\alpha = \frac{1}{4b} = 0.88 (GeV)^{-2}$ \cite{John1976}.

In 1982, Lichitenberg and his collaborators\cite{Lich1982} applied the semi-relativistic Hamiltonian (the ``Krolikowski'' type second order differential equation \cite{Krol1980,Krol1981,Todo1971}) to calculate the meson and baryon masses.

According to the lattice gauge theory, quarks are bounded by a confining potential that increases with $r$ inside a hadron, for large $r$. Here, $r$ is the separation between two quarks and the color force is a Coulomb-like potential such as $\sim 1/r$ at small $r$. In a two particle system, the simplified relativistic spin-free Hamiltonian involving only the scalar potential is described as \cite{1991,Lich1982} 
\begin{equation}
H  = \sqrt{\left( m_1 + \frac{1}{2}S(r) \right)^2 + \vec{\mathbf{p}}_{(1)}^2 } + \sqrt{\left( m_2 + \frac{1}{2}S(r) \right)^2 + \vec{\mathbf{p}}_{(2)}^2 }
\label{eq:ys}
\end{equation}
where $ \vec{\mathbf{p}}_{(k)} = -i \nabla^{(k)} $ with $k = 1,2$. Here, the scalar potential $S(r) = br$ and $b$ is related to the universal slope of the Regge trajectory. In the center of mass system, $\vec{\mathbf{p}}_{(1)} = - \vec{\mathbf{p}}_{(2)} = \vec{\mathbf{p}}$. 
Then, the semi-relativistic Hamiltonian of the system is given by
\begin{equation}
H = \sqrt{\left( m_1 + \frac{1}{2}br \right)^2 + \vec{\mathbf{p}^2 }} + \sqrt{\left( m_2 + \frac{1}{2}br  \right)^2 + \vec{\mathbf{p}^2 }}
\label{jjj:1}
\end{equation} 
With $m_1 = m_2 = m$ for the quark-antiquark system, we have     

\begin{equation}
H = 2\sqrt{\left( m + \frac{1}{2}br \right)^2 + P_r^2 + \frac{L(L+1)}{r^2} } 
\label{jj:2}
\end{equation}
where $P_r^2 = - \frac{1}{r}\frac{\partial ^2}{\partial r^2}r $. The eigenfunction $\Psi (r)$ for $H^2$ has an eigenvalue $E^2$ such that
\begin{equation}
  4\left[\left( m + \frac{1}{2}br \right)^2 - \frac{1}{r} \frac{d ^2}{d r^2}r + \frac{L(L+1)}{r^2} \right] \Psi(r) = E^2 \Psi(r) 
\label{eq:67}
\end{equation}
 G\"{u}rsey \textit{et al.} neglected the quark mass in Eq.(\ref{eq:67}) in their supersymmetric wave equation and observed that its differential equation was equivalent to a confluent hypergeometric equation having a 2-term recurrence relation between successive coefficients in its classical formal series. They obtained the eigenvalues as
\begin{equation}
E^2 = 4b( N_r + L+ 3/2)
\label{eq:100}
\end{equation}
where $N_r = 0,1,2,\cdots$ is the radial quantum number, and the Regge trajectories had slopes of $\frac{1}{4b}$ in the plots of $L$ versus $M^2$. Their theory was consistent with experimental observations \cite{1985,1988, 1991, 2011}. 

Now, we examine the changes in the eigenvalues of Eq.(\ref{eq:67}) if the quark mass is included.
Factoring out the asymptotic behaviors of the wave function $\Psi(r)$ near $r = 0$ and $r = \infty $ where $\Psi(r) = R(r) Y_L^M(\theta , \phi )= \exp\left(-\frac{b}{4}\left(r+\frac{2m}{b}\right)^2\right) r^L y(r)Y_L^M(\theta , \phi )$, the differential equation for Eq.(\ref{eq:67}) becomes
\begin{equation}
r\frac{\partial^2{y}}{\partial{r}^2} + \left( - b r^2 -2m r + 2(L+1)\right) \frac{\partial{y}}{\partial{r}} + \left( \left(\frac{E^2}{4} -
b\left( L + \frac{3}{2}\right)\right) r -2m(L+1)\right) y = 0
\label{eq:68}
\end{equation}
On comparing Eq.(\ref{eq:68})with Eq.(\ref{eq:1}), we can see that the former is a special case of the latter with $z = r $, $\mu = -b$, $\varepsilon = -2m $, $\nu = 2(L+1)$, $\omega = L+1 $, and $\Omega = \frac{E^2}{4}-b\left( L+\frac{3}{2}\right)$.

We investigate the asymptotic behavior of the radial wave function $R(r)$ in Eq.(\ref{eq:68}) as the variable $r$ approaches positive infinity. We assume that $y(r)$ is an infinite series in Eq.(\ref{eq:68}) and substitute Eq.(\ref{eq:19}) in $R(r) = \exp\left(-\frac{b}{4}\left(r+\frac{2m}{b}\right)^2\right) r^L y(r) $ 

\begin{equation}
R(r) \sim  \mathcal{A}\; r^L \left( \frac{b}{2}r^2 \right)^{-\frac{E}{8b}-\frac{1}{2}\left(L+\frac{3}{2} \right)} \exp\left(\frac{b}{4}r^2-m\left( r+\frac{m}{b} \right)\right)
\label{eq:70}
\end{equation}
In Eq.(\ref{eq:70}) if $r\rightarrow \infty $, then $\displaystyle {R(r)\rightarrow \infty }$. It is unacceptable for the wave function
$R(r)$ to be divergent as $r$ approaches infinity, from the quantum mechanical point of view. Therefore, the function $y(r)$ must be a polynomial in
Eq.(\ref{eq:68}) for the wave function $R(r)$ to be convergent, even if $r$ approaches infinity.

For polynomials of Eq.(\ref{eq:68}) around $r = 0$, we consider $b$ and $E$ to be quantized values.
The general expression of a power series of Eq.(\ref{eq:68}) about $r = 0$ for the polynomial and its algebraic equation for the determination of an accessory parameter $b$ are given by
\begin{enumerate}[1.]
\item For $N = 0$, Eq.(\ref{eq:21}) gives $B_1 = \frac{-\Omega}{2(2L+3)} = 0$ and $d_1 = A_0 d_0 = m d_0 = 0$. If we choose $d_{0} = 0$, the solution of the whole series 
vanishes. Therefore, there is no solution unless $m = 0$, in which case the solution is reduced to that of the confluent hypergeometric case where $E^{2} = 4b(L+3/2) $. 
 As we are considering the case where $m\neq 0$, we conclude that there is no solution for $N = 0$.
\item For $N \geq1$, the energy eigenvalue is determined from $B_{N+1} = 0$, giving $E^2 = 4 b \left( N + L + \frac{3}{2}\right)$ with $L = 0,1,2,\cdots$.  The quantized allowed values of $b$'s are obtained from $d_{ N+1} = 0$. Its eigenfunction is an $N$th order polynomial $y(r) = \sum_{n = 0}^{N}d_n r^n$. 
\end{enumerate}

An algebraic equation of degree $ N/2$ for the determination of $b/m^2$ for $N = $ even number has $ N/2$ real roots for a given $N$ and a characteristic equation of degree $(N+1)/2$ for $N = $ odd number has $(N+1)/2$ real roots for a given $N$.
We obtain the numeric for every real value of $ b/m^2 $ for given $N$ and $L$ to solve the algebraic equations of $ b/m^2 $.

\bgroup
\def\arraystretch{1.5}
\begin{table}[!htb]
      \centering
        \begin{tabular}{|l|l|}
\hline
$K=0$  &   $  \frac{2.17476L+1.23455N+3.1316}{N^2+0.127159N-0.0365734} $     \\   \hline
$K=1 $ & $\frac{2.07336L+1.30259(N-2)+6.42638}{(N-2)^2+0.115047(N-2)-0.0496256}$     \\ \hline
$K=2 $ & $\frac{2.01984L+1.34965(N-4)+9.57483}{(N-4)^2+0.081705(N-4)-0.0383507}$      \\ \hline
$K=3 $ & $\frac{1.9896L+1.3852(N-6)+12.6964}{(N-6)^2+0.0577076(N-6)-0.0279763} $      \\ \hline
$K=4 $ & $\frac{1.97123L+1.41335(N-8)+15.8181}{(N-8)^2+0.0416747(N-8)-0.0205087}$      \\ \hline
$K=5 $ & $\frac{1.95917L+1.43668(N-10)+18.9437}{(N-10)^2+0.0305489(N-10)-0.0151151} $   \\ \hline
$K=6 $ & $\frac{1.95079L+1.45671(N-12)+22.0738}{(N-12)^2+0.0225566(N-12)-0.0111519} $    \\ \hline
$K=7 $ & $\frac{1.94469L+1.47448(N-14)+25.2076}{(N-14)^2+0.0165878(N-14)-0.00814524}$    \\ \hline
$K=8 $ & $\frac{1.94007L+1.49074(N-16)+28.3441}{(N-16)^2+0.0119884(N-16)-0.00580511} $   \\ \hline
$K=9 $ & $\frac{1.93646L+1.50609(N-18)+31.4824}{(N-18)^2+0.00837235(N-18)-0.00396296}$   \\ \hline
$K=10$ & $\frac{1.93354L+1.5212(N-20)+34.6212}{(N-20)^2+0.00543953(N-20)-0.00247279} $   \\ \hline
\end{tabular}
\caption{\scriptsize Fit lines of $ b/m^2 $ for variables $N$ and $L$ where $N\geq 2K+1$} \label{1}
\end{table}
\egroup

In Table~\ref{1}, for $K = 0$, $b/m^2 = \frac{2.17476L+1.23455N+3.1316}{N^2+0.127159N-0.0365734} $ for $N$ and $L$ is a least-squares fit line to a list of data as a linear combination of $ b/m^2 $ of the variables $N$ and $L$. We chose 350 different smallest values of $ b/m^2 $s at $(N,L)$ for given $N$ and $L$, where $1\leq N \leq 25$ \& $0\leq L \leq N$. For example, there are 5 possible real values of $b/m^2$ for $N = L = 10$ such as 0.366018, 0.579236, 1.03967, 2.35494, and 9.45702. We chose 0.366018 as our numeric real value of $b/m^2$.

For $K = 1$, $b/m^2 = \frac{2.07336L+1.30259(N-2)+6.42638}{(N-2)^2+0.115047(N-2)-0.0496256} $ for $N$ and $L$ is a least-squares fit line to a list of data as a linear combination of $b/m^2 $ of the variables $N$ and $L$. We chose 345 different second smallest values of $b/m^2 $s at $(N,L)$ for given $N$ and $L$, where $3\leq N \leq 25$ \& $0\leq L \leq N$.

For $K = 2$, $b/m^2 = \frac{2.01984L+1.34965(N-4)+9.57483}{(N-4)^2+0.081705(N-4)-0.0383507}$ for $N$ and $L$ is a least-squares fit line to a list of data as a linear combination of $ b/m^2$ of the variables $N$ and $L$. We chose 336 different third smallest values of $ b/m^2 $s at $(N,L)$  for given $N$ and $L$, where $5\leq N \leq 25$ \& $0\leq L \leq N$.

For $K = 10$, $b/m^2 = \frac{1.93354L+1.5212(N-20)+34.6212}{(N-20)^2+0.00543953(N-20)-0.00247279}$ for $N$ and $L$ is a least-squares fit line to a list of data as a linear combination of $ b/m^2 $ of variables $N$ and $L$. We chose 120 different eleventh smallest values of $b/m^2$s at $(N,L)$ for given $N$ and $L$, where $21\leq N \leq 25$ \& $0\leq L \leq N$.

From Table~\ref{1}, we obtain the following general least-squares fit line of $ b/m^2$ of variables $N$ and $L$ for given $K$: 
\begin{equation}
\footnotesize{ b/m^2 = \left\{ \frac{\left(\frac{22}{13} - \frac{11/4}{K + 6}\right)(N - 2K) + \left(\frac{17}{9} + \frac{1/2}{ K + 7/4} \right)L + \frac{22}{7}K + \frac{13}{4}}{(N-2K)^2 + \frac{1}{K^2+8}(N-2K-1/2)}:\right. \left. \begin{matrix} 
    K\in \{0,1,2,\cdots, (N-1)/2 \} \;\mbox{for} \; \mbox{odd}\;N \\
     K\in \{0,1,2,\cdots, N/2-1 \} \; \mbox{for} \; \mbox{even}\;N \\ 
   \end{matrix} \right\}} \label{jj:1}
\end{equation} 

\begin{figure}[!htb]
\minipage{0.5\textwidth}
  \includegraphics[width = \linewidth]{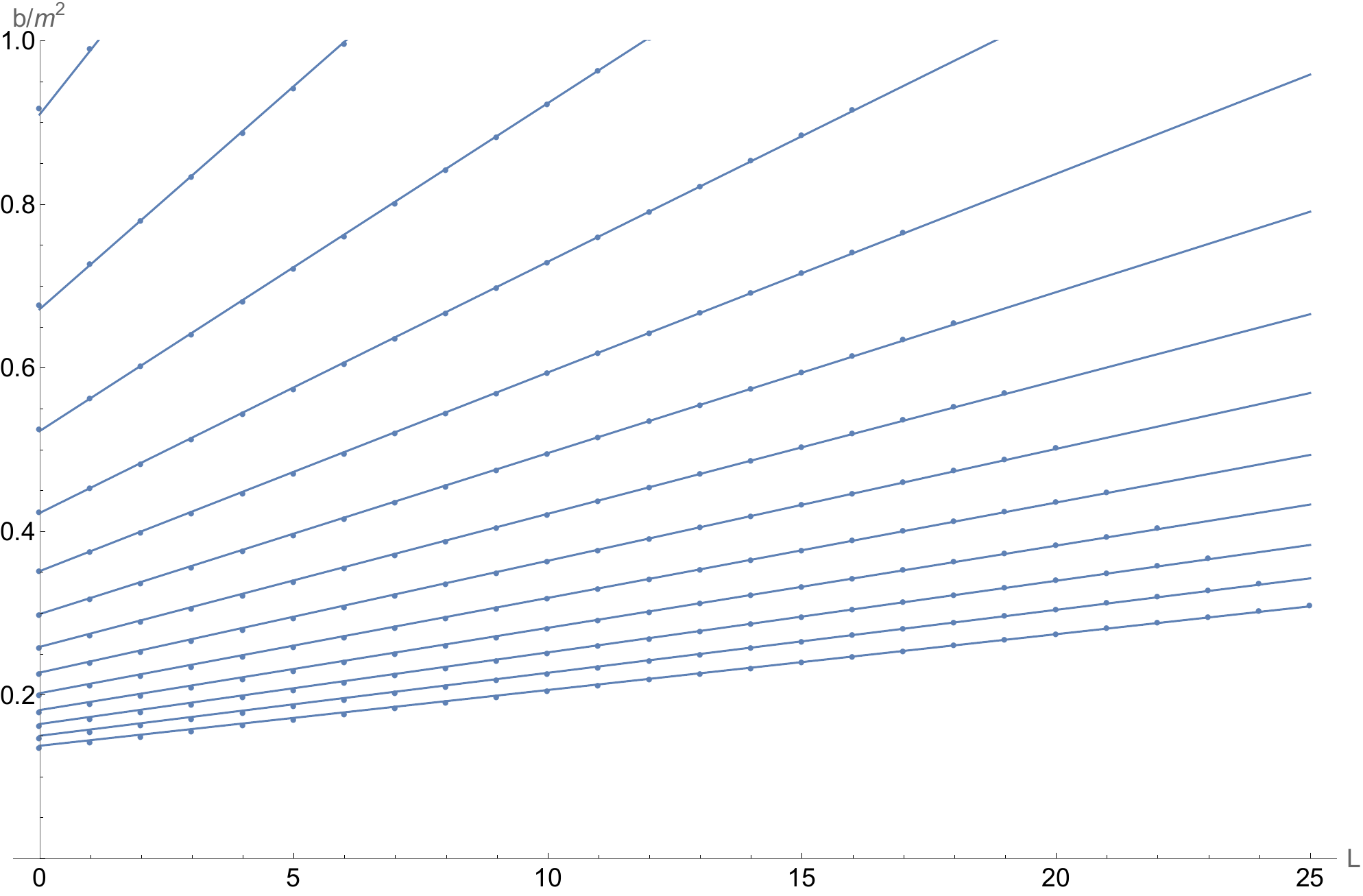}
  \caption{Fitting of $b/m^2$ given by Eq.(\ref{jj:1}) as functions of $L$ \\with a few fixed values of $N$ for $K = 4$ }
\label{A4L}
\endminipage\hfill
\minipage{0.5\textwidth}
  \includegraphics[width = \linewidth]{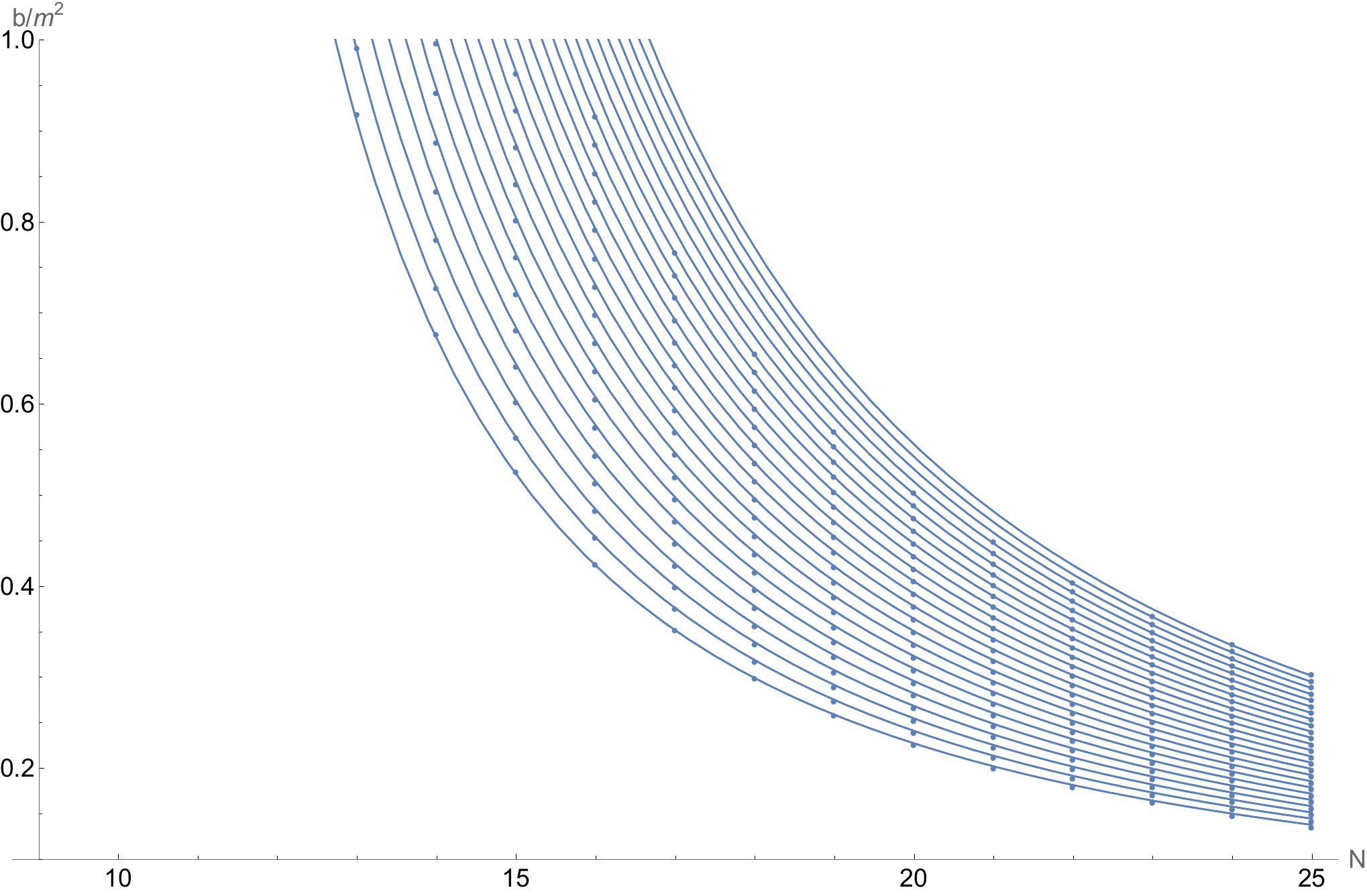}
  \caption{Fitting of $b/m^2$ given by Eq.(\ref{jj:1}) as functions of $N$ with a few fixed values of $L$ for $K = 4$}
\label{A4N}
\endminipage
\end{figure}

\begin{figure}[!htb]
\minipage{0.5\textwidth}
  \includegraphics[width = \linewidth]{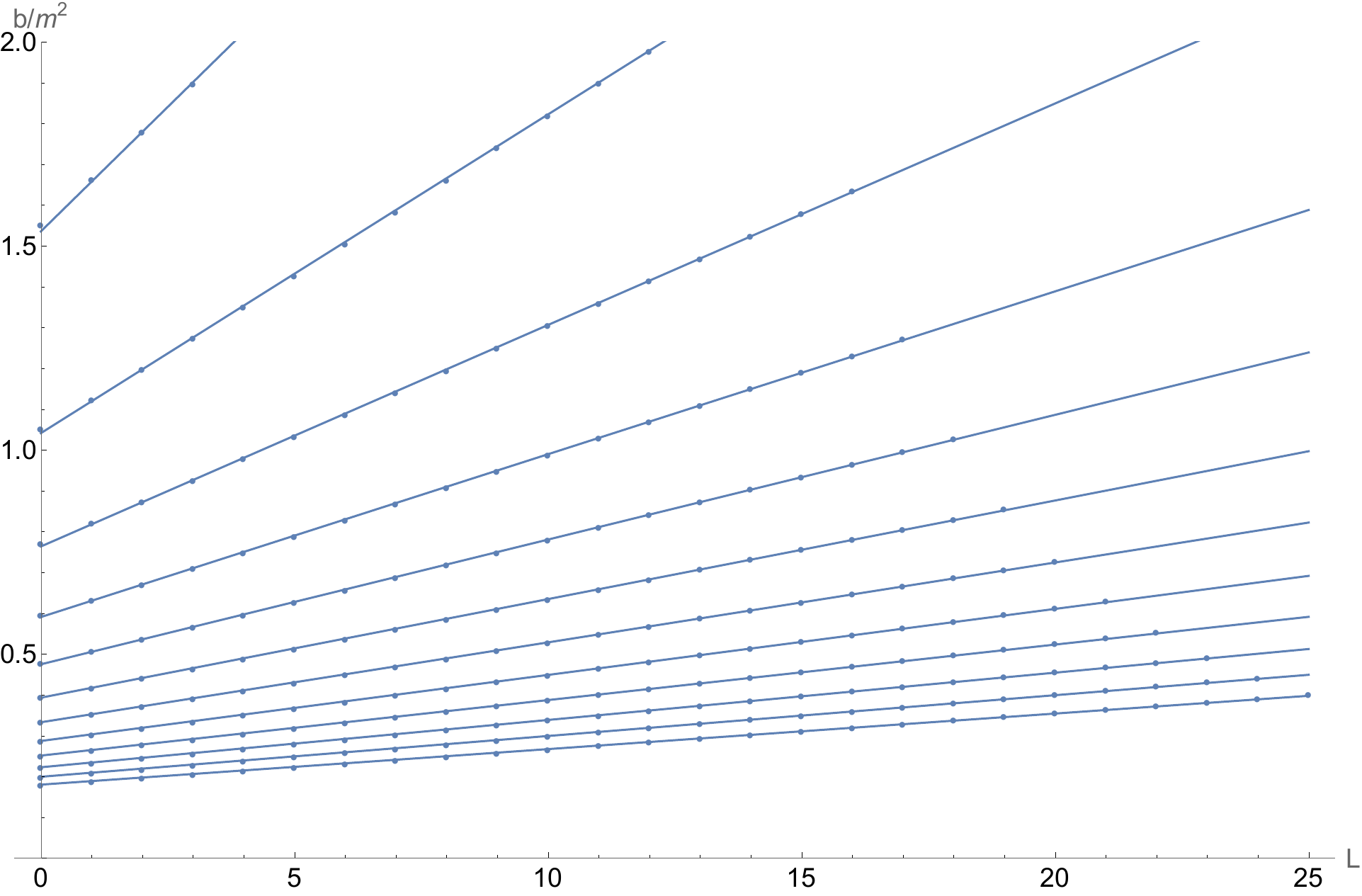}
  \caption{Fitting of $b/m^2$ given by Eq.(\ref{jj:1}) as functions of $L$ \\with a few fixed values of $N$ for $K = 5$ }
\label{A5L}
\endminipage\hfill
\minipage{0.5\textwidth}
  \includegraphics[width = \linewidth]{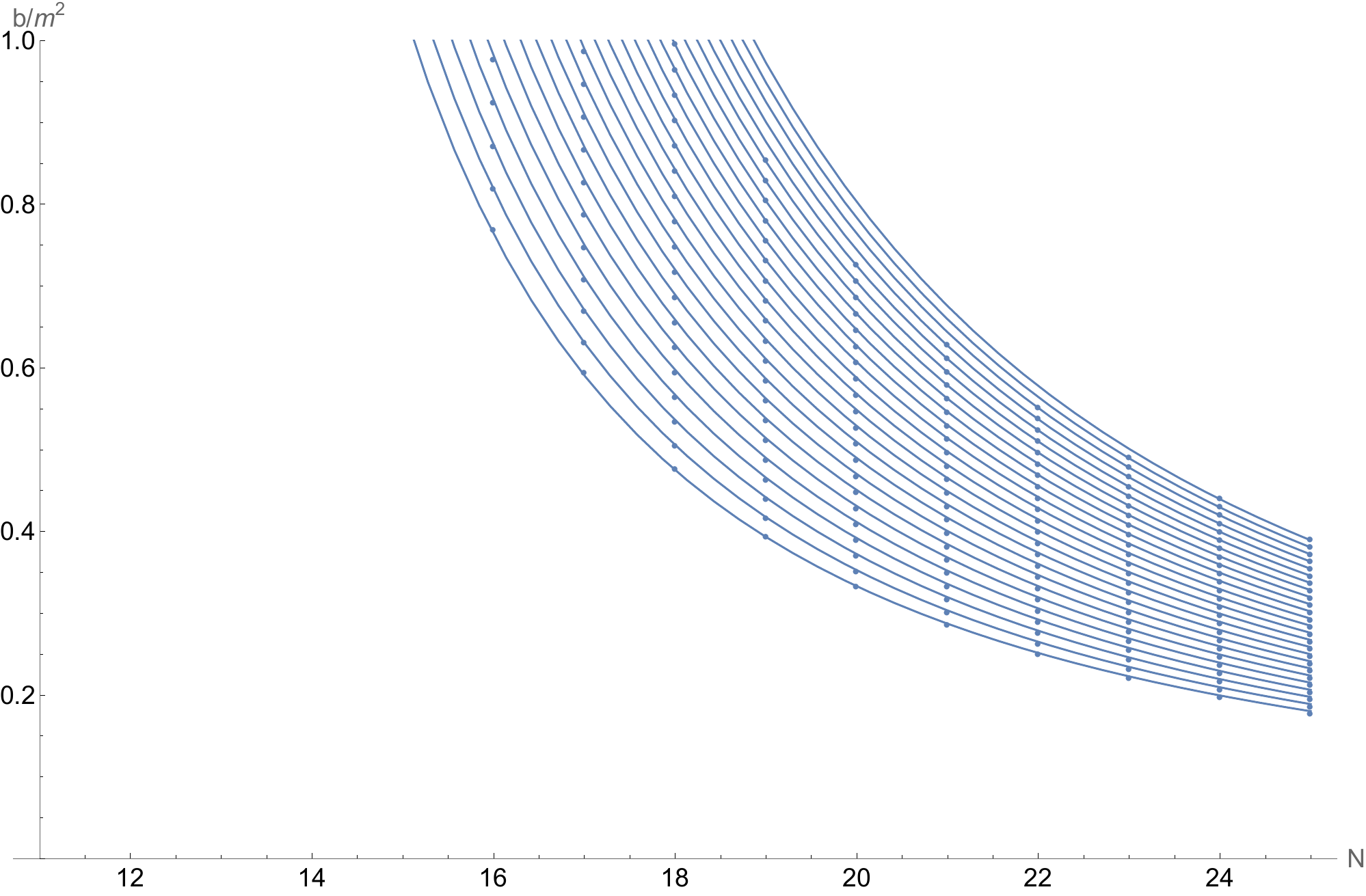}
  \caption{Fitting of $b/m^2$ given by Eq.(\ref{jj:1})as functions of $N$ with a few fixed values of $L$ for $K = 5$}
\label{A5N}
\endminipage
\end{figure}


Fig.~\ref{A4L} shows that each of the $ b/m^2 $ points is positioned on fit lines of Eq.(\ref{jj:1}) with given $N$'s for $K = 4$. The lowest fit line is for $N = 25$, the next fit is for $N = 24$, and the top line, which has the steepest slope, is for $N = 13$.
Fig.~\ref{A4N} shows that each of the $ b/m^2 $ points is positioned on fit lines of Eq.(\ref{jj:1}) with given $L$'s for $K = 4$. The lowest fit line is for $L = 0$, the next fit is for $L = 1$, and the top line is for $L = 24$.

Fig.~\ref{A5L} shows that each of the $ b/m^2 $ points is positioned on fit lines of Eq.(\ref{jj:1}) with given $N$'s for $K = 5$. The lowest fit line is for $N = 25$, the next fit is for $N = 24$, and the top line, which has the steepest slope, is for $N = 14$.
Fig.~\ref{A5N} shows that each of the $ b/m^2 $ points is positioned on fit lines of Eq.(\ref{jj:1}) with given $L$'s for $K = 5$. The lowest fit line is for $L = 0$, the next fit is for $L = 1$, and the top line is for $L = 24$.

By substituting Eq.(\ref{jj:1}) in {\footnotesize $E^2 = 4 b \left( N + L + \frac{3}{2}\right) $}, we obtain the experimental
fit to the set of eigenvalues $E^2$ as given below: 
\begin{equation}
\footnotesize{E^2 = \left\{ 4 \left(\frac{\left(\frac{22}{13} - \frac{11/4}{K + 6}\right)(N - 2K) + \left(\frac{17}{9} + \frac{1/2}{ K + 7/4} \right)L + \frac{22}{7}K + \frac{13}{4}}{(N - 2K)^2 + \frac{1}{K^2 + 8}(N - 2K - 1/2)} \right) \left( N + L + \frac{3}{2}\right)m^2:\right. \left. \begin{matrix} 
    K\in \{0,1,2,\cdots, (N-1)/2 \} \;\mbox{for} \; \mbox{odd}\;N \\
     K\in \{0,1,2,\cdots, N/2-1 \} \; \mbox{for} \; \mbox{even}\;N \\ 
   \end{matrix} \right\}} \label{Bag}
\end{equation}

The highest real roots of $b/m^2$ for a given $(N,L)$ in Eq.(\ref{jj:1}) yield the lowest energy eigenvalues satisfying $ E_0^2 < E_2^2 < E_4^2 < \cdots $ and $ E_1^2 < E_3^2 < E_5^2 < \cdots $ where, $E_{N}^2$ is the $N^{th}$ excited eigenvalue for a given $L$. This means that we collect each $E^2$ of Eq.(\ref{Bag}) where, $(N,K) = (N,(N-1)/2) $ for odd $N$, and it gives the following fit for $E^2$: 
\be
  E_{N,L}^2 \approx \frac{32}{5}\left( N + \frac{6}{5}L+\frac{5}{3}\right) \left( N + L + \frac{3}{2}\right)m^2. \label{energyF}
 \ee
Similarly, for even $N$, we collect each $E^2$ of Eq.(\ref{Bag}) where $(N,K) = (N,N/2-1) $, and we have the following fit:
\be
   E_{N,L}^2 \approx \frac{8}{5}\left( N + \frac{6}{5}L + \frac{5}{3}\right) \left( N + L + \frac{3}{2}\right)m^2 \label{energyG}
 \ee

The mass spectrum that is approximately given by Eq.(\ref{energyF}) and Eq.(\ref{energyG}) cannot be linear in $N$ unlike in the case where $m = 0$ as given by Eq.(\ref{eq:100}).
$b$ is determined by other parameters, which in turn introduces an additional dependence of $E^2$ on $N$ and $L$ through that of $b$.

Many researchers observed that excited mesons, including baryons, appear to have an elongated bilocal linear structure held by a gluon flux tube with a scalar linear potential for high rotational excitation at large separation.
G\"{u}rsey and collaborators obtained the semi-relativistic Hamiltonian for the $q-\bar{q}$ system neglecting the small mass of quarks\cite{1991}, as suggested by Lichitenberg \textit{et al.}\cite{Lich1982}. In this system, the QCD forces were flavor-independent, the strong-coupling potential (similar to the Coulomb potential) was negligible, and the confining part of the QCD potential was spin-independent.
The wave equation determined by neglecting the quark mass was equivalent to a confluent hypergeometric differential equation in which the recursive relation involved two terms in its power series expansion. They obtained a one-mass formula with a universal Regge slope for mesons, which was consistent with experimental observations.

However, if we include the quark mass into their spin-free Hamiltonian involving only a scalar potential, we obtain a modified form of the BCH equation. The quark mass triggers a change from the hypergeometric to Heun's type of singularity. Its recurrence relation consists of three terms in the formal series solution. For the 2-term recursive relation in a formal series, only a single quantized parameter ($E^2$) is sufficient to obtain a polynomial solution; however, in the 3-term case, we need two parameters to construct a polynomial, namely, a tension $b$ and an energy $E^2$ that must be quantized.

On comparing G\"{u}rsey's mass formula with Eq.(\ref{energyF}) and Eq.(\ref{energyG}), we observe that if $m$ is non-zero, the tension $b$ is determined by $m$, $N$, and $L$, and the linearity of the spectrum disappears i.e., it exhibits a non-linear behavior and the model is inconsistent with hadron spectrum phenomenology.  Therefore, to fit our theory to the data, the current quark mass must be zero. If $m = 0$, $b$ is not determined by $m$ and has an intrinsic scale by itself. Its spectrum shows linearity by its scale.
Thus, G\"{u}rsey's assumption of negligible current quark mass in his mass formula is valid.
We infer that chiral symmetry is induced by color confinement and suggest that the ``chiral symmetry must be related to confinement dynamics'' by understanding the role of the quark mass to change the singularity structure in a simple model.

 \appendix
\section{Biconfluent Heun equation}

\begin{equation}
z \frac{d^2{y}}{d{z}^2} + \left( \mu z^2 + \varepsilon z + \nu \right) \frac{d{y}}{d{z}} + \left( \Omega z + \varepsilon \omega \right) y = 0
\label{eq:1}
\end{equation}
Eq.(\ref{eq:1}) is a modified form of the BCH equation where $\mu$, $\varepsilon$, $\nu $, $\Omega$, and $\omega$ are real or imaginary
parameters. It has a regular singularity at the origin and an irregular singularity at infinity of rank 2 \cite{Bato1977,Caru2014,Deca1978,Haut1969,Ronv1995,Urwi1975}.   
Recently, this equation was applied to the Schr$\ddot{\mbox{o}}$dinger equation to determine the second Exton potential and power potentials \cite{Ishk2019,Karw2015}. 

For the Heun equation having 4 regular singular points, the recurrence relation, in its Frobenius solution, involves 3 terms. The Heun equation generalizes all the well-known equations of a spheroidal wave, Lame, Mathieu, and hypergeometric type.
Until now, its series solutions in which the coefficients are given fully and clearly, are unknown owing to the complex mathematical computations, and their numerical calculations are still ambiguous. For definite or contour integrals of the Heun equation, no analytical solutions have been constructed as yet.

Similar to the derivation of confluent hypergeometric equations from hypergeometric equations, 4 confluent types of Heun equations can be derived by merging two or more regular singularities to consider an irregular singularity in the Heun equation. These types include (1) confluent Heun (two regular and one irregular singularities), (2) doubly confluent Heun (two irregular singularities), (3) BCH (one regular and one irregular singularities), (4) triconfluent Heun equations (one irregular singularity).

$y(z)$ has a series expansion of the form around the origin given by
\begin{equation}
y(z) = \sum_{n = 0}^{\infty } d_n z^{n }
\label{eq:2}
\end{equation}
  Putting Eq.(\ref{eq:2}) into Eq.(\ref{eq:1}) we obtain
\begin{equation}
d_{n+1} = A_n \;d_n + B_n \;d_{n-1} \hspace{1cm};n\geq 1
\label{eq:3}
\end{equation}
 where $A_n = -\frac{\varepsilon (n + \omega )}{(n+1 )(n+\nu )}$, $B_n = -\frac{\Omega + \mu (n-1 )}{(n+1 )(n+\nu )}$, and $d_1 = A_0 \;d_0$.

\section{Asymptotic behavior of a modified BCH function}
We now investigate the function $y(z)$ as $z$ approaches infinity. Assume that $y(z)$ is an infinite series and its series expansion of
Eq.(\ref{eq:2}) is given by

\begin{eqnarray}
 y(z)& = & \sum_{m=0}^{\infty}y_m(\tilde{z}) \nonumber\\
&=& \sum_{i_0=0}^{\infty } \frac{(\frac{\Omega }{2\mu })_{i_0}}{(1)_{i_0}(\gamma)_{i_0}}\tilde{z}^{i_0} + \tilde{\varepsilon } \sum_{i_0=0}^{\infty }
\frac{(i_0+\frac{\omega }{2})}{(i_0+\frac{1}{2})(i_0-\frac{1}{2}+\gamma )} \frac{(\frac{\Omega }{2\mu })_{i_0}}{(1)_{i_0}(\gamma)_{i_0}}\nonumber\\
&&\times \sum_{i_1=i_0}^{\infty }  \frac{(\frac{\Omega }{2\mu }+\frac{1}{2})_{i_1}(\frac{3}{2})_{i_0}(\gamma +\frac{1}{2})_{i_0}}{(\frac{\Omega }{2\mu
}+\frac{1}{2})_{i_0}(\frac{3}{2})_{i_1}(\gamma +\frac{1}{2})_{i_1}} \tilde{z}^{i_1}
+ \sum_{n=2}^{\infty } \tilde{\varepsilon }^n \sum_{i_0=0}^{\infty } \frac{(i_0+\frac{\omega }{2})}{(i_0+\frac{1}{2})(i_0-\frac{1}{2}+\gamma)}
\frac{(\frac{\Omega }{2\mu })_{i_0}}{(1)_{i_0}(\gamma )_{i_0}} \nonumber\\
&&\times \prod _{k=1}^{n-1} \Bigg\{ \sum_{i_k=i_{k-1}}^{\infty } \frac{(i_k+\frac{\omega
}{2}+\frac{k}{2})}{(i_k+\frac{1}{2}+\frac{k}{2})(i_k-\frac{1}{2}+\gamma + \frac{k}{2})} \frac{(\frac{\Omega }{2\mu
}+\frac{k}{2})_{i_k}(1+\frac{k}{2})_{i_{k-1}}(\frac{k}{2}+\gamma )_{i_{k-1}}}{(\frac{\Omega }{2\mu
}+\frac{k}{2})_{i_{k-1}}(1+\frac{k}{2})_{i_k}(\frac{k}{2}+\gamma )_{i_k}}\Bigg\} \nonumber\\
&&\times \sum_{i_n= i_{n-1}}^{\infty } \frac{(\frac{\Omega }{2\mu }+\frac{n}{2})_{i_n}(1+\frac{n}{2})_{i_{n-1}}(\frac{n}{2}+\gamma
)_{i_{n-1}}}{(\frac{\Omega }{2\mu }+\frac{n}{2})_{i_{n-1}}(1+\frac{n}{2})_{i_n}(\frac{n}{2}+\gamma )_{i_n}} \tilde{z}^{i_n}
\label{eq:6}
\end{eqnarray}
where
\begin{equation}
\begin{cases} \tilde{z} = -\frac{1}{2}\mu z^2 \cr
\tilde{\varepsilon } = -\frac{1}{2}\varepsilon  z\cr
\gamma = \frac{1}{2}(1+\nu )
\end{cases}
\nonumber
\end{equation}
In the above equations, $y_m(\tilde{z})$ is an $m$-tuple series; $y_0(\tilde{z})$ is a single series, $y_1(\tilde{z})$ is a double series, $y_2(\tilde{z})$ is a triple series, and so on.
In this study, the Pochhammer symbol $(x)_n$ was used to represent the rising factorial $(x)_n = \frac{\Gamma (x+n)}{\Gamma (x)}$.
More precisely, the sequence $ d_{n} $ is a combination of $A_n$ and $B_n$ terms in Eq.(\ref{eq:3}). Eq.(\ref{eq:6}) is obtained by considering $A_n$ in the sequence $d_{n}$ as the leading term in a series $\sum_{n=0}^{\infty } d_n z^{n }$. We observe that the terms of the sequence $d_{n}$ have a zero term of $A_n's$ for a sub-power series $y_0(\tilde{z})$, one term of $A_n's$ for the sub-power series $y_1(\tilde{z})$, two terms of $A_n's$ for a $y_2(\tilde{z})$, three terms of $A_n's$ for a $y_3(\tilde{z})$, and so on. 

There is a generalized hypergeometric function that is given by

\begin{eqnarray}
L_j &=& \sum_{i_j= i_{j-1}}^{\infty } \frac{\left(\frac{\Omega }{2\mu }+\frac{j}{2}\right)_{i_j}(1+\frac{j}{2})_{i_{j-1}}(\frac{j}{2}+\gamma
)_{i_{j-1}}}{\left(\frac{\Omega }{2\mu }+\frac{j}{2}\right)_{i_{j-1}}(1+\frac{j}{2})_{i_j}(\frac{j}{2}+\gamma )_{i_j}} \tilde{z}^{i_j}\nonumber\\
&=& \tilde{z}^{i_{j-1}}
\sum_{l=0}^{\infty } \frac{B(i_{j-1}+\frac{j}{2},l+1) B(i_{j-1}-1+\gamma +\frac{j}{2},l+1)\left( \frac{\Omega }{2\mu
}+\frac{j}{2}+i_{j-1}\right)_l}{(i_{j-1}+\frac{j}{2})^{-1}(i_{j-1}-1+\gamma +\frac{j}{2})^{-1}(1)_l \;l!} \tilde{z}^l \hspace{1cm}\label{eq:7}
\end{eqnarray}
where $j = 1,2,3,\cdots$. Eq.(\ref{eq:7}) is derived by shifting an index summation $ \sum_{i_j = i_{j-1}}^{\infty } \rightarrow  \sum_{i_j= 0}^{\infty } $ and then the formula $B(x,y) = \frac{\Gamma (x)\Gamma (y)}{\Gamma (x+y)}$ is utilized.

By using the integral form of the beta function,
\begin{subequations}
\begin{equation}
B\left(i_{j-1}+\frac{j}{2}, l+1\right) = \int_{0}^{1} dt_j\;t_j^{i_{j-1}+\frac{j}{2}-1} (1-t_j)^l
\label{eq:8a}
\end{equation}
\begin{equation}
 B\left(i_{j-1}+\gamma -1+\frac{j}{2}, l+1\right) = \int_{0}^{1} du_j\;u_j^{i_{j-1}+\gamma -2+\frac{j}{2}} (1-u_j)^l
 \label{eq:8b}
\end{equation}
\end{subequations}

Substituting Eq.(\ref{eq:8a}) and Eq.(\ref{eq:8b}) into Eq.(\ref{eq:7}), and dividing the new Eq.(\ref{eq:7}) by $(i_{j-1}+\frac{j}{2})(i_{j-1}-1+\gamma +\frac{j}{2})$ 
we obtain
\begin{footnotesize}
\begin{eqnarray}
I_j &=& \frac{1}{(i_{j-1}+\frac{j}{2})(i_{j-1}-1+\gamma +\frac{j}{2})}
 \sum_{i_j= i_{j-1}}^{\infty } \frac{\left(\frac{\Omega }{2\mu }+\frac{j}{2}\right)_{i_j}(1+\frac{j}{2})_{i_{j-1}}(\frac{j}{2}+\gamma
 )_{i_{j-1}}}{\left(\frac{\Omega }{2\mu }+\frac{j}{2}\right)_{i_{j-1}}(1+\frac{j}{2})_{i_j}(\frac{j}{2}+\gamma )_{i_j}} \tilde{z}^{i_j}\nonumber\\
&=&  \int_{0}^{1} dt_j\;t_j^{\frac{j}{2}-1} \int_{0}^{1} du_j\;u_j^{\gamma -2+\frac{j}{2}} (\tilde{z} t_j u_j)^{i_{j-1}}
 \sum_{l=0}^{\infty } \frac{\left( \frac{\Omega }{2\mu }+\frac{j}{2}+i_{j-1}\right)_l}{(1)_l} \frac{\big( \tilde{z}(1-t_j)(1-u_j)\big)
 ^l}{l!}\nonumber\\
&=&  \lim_{\epsilon\rightarrow 1}\int_{0}^{\epsilon} dt_j\;t_j^{\frac{j}{2}-1} \int_{0}^{\epsilon} du_j\;u_j^{\gamma -2+\frac{j}{2}} (\tilde{z} t_j
u_j)^{i_{j-1}}
 M\left( \frac{\Omega }{2\mu }+\frac{j}{2}+i_{j-1}, 1,  \tilde{z}(1-t_j)(1-u_j)\right) \label{eq:9}
\end{eqnarray}
\end{footnotesize}
where, $M(a,b,z) = \sum_{n=0}^{\infty } \frac{(a)_n}{(b)_n n!} z^n$ is a Kummer function of the first kind.
The asymptotic behavior of Kummer's solution as the real part of $z$ approaches positive infinity is $M(a,b,z) \sim  \frac{\Gamma(b)}{\Gamma(a)} z^{a-b}
\exp(z)$. The asymptotic function of Kummer's solution in Eq.(\ref{eq:9}) is expressed as
\begin{equation}
M\left( \frac{\Omega }{2\mu }+\frac{j}{2}+i_{j-1}, 1,  \tilde{z}(1-t_j)(1-u_j)\right) \sim
\frac{\left(\tilde{z}(1-t_j)(1-u_j)\right)^{\frac{\Omega}{2\mu}-1+\frac{j}{2}+i_{j-1}}}{\Gamma(\frac{\Omega}{2\mu}+\frac{j}{2}+i_{j-1})}
\exp\left(\tilde{z}(1-t_j)(1-u_j)\right) \label{eq:10}
\end{equation}
Putting Eq.(\ref{eq:10}) into Eq.(\ref{eq:9}),
\begin{footnotesize}
 \begin{eqnarray}
I_j &\sim & \frac{\tilde{z}^{\frac{\Omega}{2\mu}-1+\frac{j}{2}+2i_{j-1}}}{\Gamma(\frac{\Omega}{2\mu}+\frac{j}{2}+i_{j-1})}   \lim_{\epsilon\rightarrow
1}\int_{0}^{\epsilon} dt_j\;t_j^{\frac{j}{2}-1+i_{j-1}}(1-t_j)^{\frac{\Omega}{2\mu}-1+\frac{j}{2}+i_{j-1}} \int_{0}^{\epsilon} du_j\;u_j^{\gamma
-2+\frac{j}{2}+i_{j-1}}(1-u_j)^{\frac{\Omega}{2\mu}-1+\frac{j}{2}+i_{j-1}}\nonumber\\
&&\times \sum_{k=0}^{\infty}\frac{\left(\tilde{z}(1-t_j)(1-u_j)\right)^k}{k!} \nonumber\\
&=& \frac{\tilde{z}^{\frac{\Omega}{2\mu}-1+\frac{j}{2}+2i_{j-1}}}{\Gamma(\frac{\Omega}{2\mu}+\frac{j}{2}+i_{j-1})}
\sum_{k=0}^{\infty}\frac{\tilde{z}^k}{k!}  \lim_{\epsilon\rightarrow 1}\int_{0}^{\epsilon}
dt_j\;t_j^{\frac{j}{2}-1+i_{j-1}}(1-t_j)^{\frac{\Omega}{2\mu}-1+\frac{j}{2}+i_{j-1}+k}  \int_{0}^{\epsilon} du_j\;u_j^{\gamma -2+\frac{j}{2}+i_{j-1}}(1-u_j)^{\frac{\Omega}{2\mu}-1+\frac{j}{2}+i_{j-1}+k} \nonumber\\
&=& \frac{\tilde{z}^{\frac{\Omega}{2\mu}-1+\frac{j}{2}+2i_{j-1}}}{\Gamma(\frac{\Omega}{2\mu}+\frac{j}{2}+i_{j-1})}
\sum_{k=0}^{\infty}\frac{\tilde{z}^k}{k!} \int_{0}^{1} dt_j\;t_j^{\frac{j}{2}-1+i_{j-1}}(1-t_j)^{\frac{\Omega}{2\mu}-1+\frac{j}{2}+i_{j-1}+k}  \int_{0}^{1} du_j\;u_j^{\gamma -2+\frac{j}{2}+i_{j-1}}(1-u_j)^{\frac{\Omega}{2\mu}-1+\frac{j}{2}+i_{j-1}+k} \nonumber\\
&=& \frac{\tilde{z}^{\frac{\Omega}{2\mu}-1+\frac{j}{2}+i_{j-1}}}{\Gamma(\frac{\Omega}{2\mu}+\frac{j}{2}+i_{j-1})} \sum_{k=0}^{\infty}
B\left(\frac{j}{2}+i_{j-1}, \frac{\Omega}{2\mu}+\frac{j}{2}+i_{j-1}+k\right)  B \left( \gamma -1+\frac{j}{2}+i_{j-1}, \frac{\Omega}{2\mu} +\frac{j}{2}+i_{j-1}+k\right) \frac{\tilde{z}^k}{k!} \label{eq:11}
\end{eqnarray}
\end{footnotesize}
We know that 
\begin{equation}
\sum_{k=0}^{\infty} \frac{B \left( a,b+k \right)B \left( c,b+k  \right)}{k!}z^k = B \left ( a,b \right) B \left( c,b \right)\; \pFq[4]{2}{2}{b,b}{a+b,c+b}{z}   \label{eq:12}
\end{equation}
Applying Eq.(\ref{eq:12})to Eq.(\ref{eq:11}), we obtain the following expression for $I_j$:
\begin{eqnarray}
I_j &\sim &  \frac{\tilde{z}^{\frac{\Omega}{2\mu}-1+\frac{j}{2}+2i_{j-1}}}{\Gamma(\frac{\Omega}{2\mu}+\frac{j}{2}+i_{j-1})}  B\left(\frac{j}{2}+i_{j-1},
\frac{\Omega}{2\mu}+\frac{j}{2}+i_{j-1} \right)  B\left( \gamma -1+\frac{j}{2}+i_{j-1}, \frac{\Omega}{2\mu} +\frac{j}{2}+i_{j-1} \right) \nonumber\\
&&\times\; \pFq[4]{2}{2}{\frac{\Omega}{2\mu}+\frac{j}{2}+i_{j-1},\frac{\Omega}{2\mu}+\frac{j}{2}+i_{j-1}}{\frac{\Omega}{2\mu}+j +2i_{j-1},\frac{\Omega}{2\mu}+\gamma-1 + j + 2i_{j-1}}{ \tilde{z}}  \label{eq:13}
\end{eqnarray}
The asymptotic behavior of a $_2F_2$ function for large $|z|$ is given by \cite{Olve2010}  
\begin{equation}
\pFq[4]{2}{2}{a_1,a_2}{b_1,b_2}{z}\sim \frac{\Gamma(b_1) \Gamma(b_2)}{\Gamma(a_1)\Gamma(a_2)}z^{a_1+a_2-b_1-b_2} e^{z}\label{qq:14}
\end{equation}
Substituting Eq.(\ref{qq:14}) into Eq.(\ref{eq:13}), we have the following asymptotic function for $I_j$:
\begin{equation}
I_j \sim  \frac{\Gamma\left(\frac{j}{2}\right)\Gamma\left(\gamma-1+\frac{j}{2}\right)}{\Gamma\left(\frac{\Omega}{2\mu}+\frac{j}{2}\right)}
\frac{\left( \frac{j}{2}\right)_{i_{j-1}} \left( \gamma-1+\frac{j}{2}\right)_{i_{j-1}}}{\left(\frac{\Omega}{2\mu}+ \frac{j}{2}\right)_{i_{j-1}}}
\tilde{z}^{\frac{\Omega}{2\mu}-\gamma - \frac{j}{2}} e^{\tilde{z}}  \label{eq:14}
\end{equation}
The asymptotic behavior of $y_0(z)$ (Kummer function of the first kind) for large $|z|$ is given by
\begin{equation}
y_0 (\tilde{z})\sim \frac{\Gamma\left( \gamma\right)}{\Gamma\left( \frac{\Omega}{2\mu}\right)}z^{\frac{\Omega}{2\mu}-\gamma} e^{\tilde{z}}\label{eq:15}
\end{equation}
Putting $j = 1$ in Eq.(\ref{eq:14}) and incorporating the new Eq.(\ref{eq:14}) into the expression for $y_1(z)$ in Eq.(\ref{eq:6}), we obtain
\begin{eqnarray}
y_1(\tilde{z})  &=& \tilde{\varepsilon } \sum_{i_0=0}^{\infty }
\frac{(i_0+\frac{\omega }{2})}{(i_0+\frac{1}{2})(i_0-\frac{1}{2}+\gamma )} \frac{(\frac{\Omega }{2\mu })_{i_0}}{(1)_{i_0}(\gamma)_{i_0}} 
 \sum_{i_1=i_0}^{\infty }  \frac{(\frac{\Omega }{2\mu }+\frac{1}{2})_{i_1}(\frac{3}{2})_{i_0}(\gamma +\frac{1}{2})_{i_0}}{(\frac{\Omega }{2\mu
}+\frac{1}{2})_{i_0}(\frac{3}{2})_{i_1}(\gamma +\frac{1}{2})_{i_1}} \tilde{z}^{i_1} \nonumber\\
&\sim&   \left( \frac{-\varepsilon }{\sqrt{-2\mu}} \right)
\sum_{i_0=0}^{\infty }   \frac{  \left(i_0+\frac{\omega }{2}\right)\Gamma\left( i_0+\frac{1}{2}\right)\Gamma\left( i_0+\gamma-\frac{1}{2}\right)}
{\Gamma\left(i_0+ \frac{\Omega }{2\mu }+\frac{1}{2}\right)}
\frac{\left(\frac{\Omega }{2\mu }\right)_{i_0} }{\left(1\right)_{i_0}\left(\gamma\right)_{i_0} }
\tilde{z}^{\frac{\Omega}{2\mu}-\gamma}e^{\tilde{z}}\label{eq:16}
\end{eqnarray}
Putting $j = 2$ in Eq.(\ref{eq:14}) and incorporating the new Eq.(\ref{eq:14}) into the expression for $y_2(z)$ in Eq.(\ref{eq:6}), we obtain
\begin{eqnarray}
y_2(\tilde{z}) &=& \tilde{\varepsilon }^2 \sum_{i_0=0}^{\infty }
\frac{(i_0+\frac{\omega }{2})}{(i_0+\frac{1}{2})(i_0-\frac{1}{2}+\gamma )} \frac{(\frac{\Omega }{2\mu })_{i_0}}{(1)_{i_0}(\gamma)_{i_0}} 
 \sum_{i_1=i_0}^{\infty }  \frac{(i_1+\frac{\omega }{2}+\frac{1}{2})}{(i_1+1)(i_1 +\gamma )} \frac{(\frac{\Omega }{2\mu }+\frac{1}{2})_{i_1}(\frac{3}{2})_{i_0}(\gamma +\frac{1}{2})_{i_0}}{(\frac{\Omega }{2\mu}+\frac{1}{2})_{i_0}(\frac{3}{2})_{i_1}(\gamma +\frac{1}{2})_{i_1}} \nonumber\\
&&\times \sum_{i_2=i_1}^{\infty } \frac{(\frac{\Omega }{2\mu }+1)_{i_2}(2)_{i_1}(\gamma +1)_{i_1}}{(\frac{\Omega }{2\mu}+1)_{i_1}(2)_{i_2}(\gamma +1)_{i_2}} \tilde{z}^{i_2} \nonumber\\
&\sim &  \left( \frac{-\varepsilon }{\sqrt{-2\mu}} \right)^2
\sum_{i_0=0}^{\infty } \frac{\left( i_0 +\frac{\omega}{2} \right)}{\left(i_0 +\frac{1}{2} \right)\left(i_0 + \gamma -\frac{1}{2} \right) }
\frac{\left(\frac{\Omega }{2\mu}\right)_{i_0}}{\left(1\right)_{i_0}\left(\gamma\right)_{i_0}} \nonumber\\
&&\times \sum_{i_1=i_0}^{\infty } \frac{\left( i_1 +\frac{1}{2}+\frac{\omega}{2 } \right) \Gamma\left( i_1+1 \right) \Gamma\left( i_1+\gamma \right)}{
\Gamma\left( i_1 +\frac{\Omega}{2\mu}+1 \right)} \frac{\left( \frac{\Omega}{2\mu }+ \frac{1}{2}\right)_{i_1} \left(  \frac{3}{2}\right)_{i_0}\left(
\gamma+ \frac{1}{2}\right)_{i_0}}{\left( \frac{\Omega}{2\mu }+ \frac{1}{2}\right)_{i_0} \left(  \frac{3}{2}\right)_{i_1}\left( \gamma+
\frac{1}{2}\right)_{i_1}}
 \tilde{z}^{\frac{\Omega}{2\mu}-\gamma}e^{\tilde{z}}   \label{eq:17}
\end{eqnarray}
Similarly, putting $j = 3$ in Eq.(\ref{eq:14}) and incorporating the new Eq.(\ref{eq:14}) into the expression for $y_3(z)$ in Eq.(\ref{eq:6}), we obtain
\begin{footnotesize}
\begin{eqnarray}
y_3(\tilde{z}) &\sim &  \left( \frac{-\varepsilon }{\sqrt{-2\mu}} \right)^3
\sum_{i_0=0}^{\infty } \frac{\left( i_0 +\frac{\omega}{2} \right)}{\left(i_0 +\frac{1}{2} \right)\left(i_0 + \gamma -\frac{1}{2} \right) }
\frac{\left(\frac{\Omega }{2\mu}\right)_{i_0}}{\left(1\right)_{i_0}\left(\gamma\right)_{i_0}} \nonumber\\
&&\times \sum_{i_1=i_0}^{\infty } \frac{\left( i_1 +\frac{1}{2}+\frac{\omega}{2 } \right)}{
 \left( i_1+1 \right)  \left( i_1+\gamma \right)} \frac{\left( \frac{\Omega}{2\mu }+ \frac{1}{2}\right)_{i_1} \left(  \frac{3}{2}\right)_{i_0}\left(
\gamma+ \frac{1}{2}\right)_{i_0}}{\left( \frac{\Omega}{2\mu }+ \frac{1}{2}\right)_{i_0} \left(  \frac{3}{2}\right)_{i_1}\left( \gamma+
\frac{1}{2}\right)_{i_1}}\nonumber\\
&&\times \sum_{i_2=i_1}^{\infty } \frac{\left( i_2 +1 + \frac{\omega}{2} \right) \Gamma\left( i_2+\frac{3}{2} \right) \Gamma\left( i_2+\gamma +\frac{1}{2} \right)}{\Gamma\left( i_2 +\frac{\Omega}{2\mu}+\frac{3}{2} \right)} \frac{\left( \frac{\Omega}{2\mu}+ 1\right)_{i_2} \left(  2\right)_{i_1}\left(\gamma+ 1\right)_{i_1}}{\left( \frac{\Omega}{2\mu}+ 1\right)_{i_1} \left( 2\right)_{i_2}\left( \gamma+ 1\right)_{i_2}}
 \tilde{z}^{\frac{\Omega}{2\mu}-\gamma}e^{\tilde{z}}   \label{eq:18}
\end{eqnarray}
\end{footnotesize}
By repeating this process for all the higher terms of the asymptotic functions of the sub-summation $y_m(\tilde{z})$ terms where $m \geq 4$, we obtain every asymptotic form of the $y_m(\tilde{z})$ terms.
As we substitute Eq.(\ref{eq:15}), Eq.(\ref{eq:16}), Eq.(\ref{eq:17}), and Eq.(\ref{eq:18}), and include all the asymptotic forms of the $y_m(\tilde{z})$ terms where $m \geq 4$ in Eq.(\ref{eq:6}), we obtain the following asymptotic behavior of $y(z)$:
\begin{equation}
y(z)\sim  \mathcal{A}\; \tilde{z}^{\frac{\Omega}{2\mu}-\gamma}\exp(\tilde{z})  \label{eq:19}
\end{equation} 
$\mathcal{A}$ is just a fixed constant that is dependent on $\mu$, $\varepsilon$, $\nu $, $\Omega$, and $\omega$ (real or imaginary parameters).
 
\bibliographystyle{model1a-num-names}
\bibliography{<your-bib-database>}
  
\end{document}